\documentclass[conference]{IEEEtran}
\IEEEoverridecommandlockouts
\usepackage{cite}
\usepackage{amsmath,amssymb,amsfonts}
\usepackage{algorithmic}
\usepackage{graphicx}
\usepackage{textcomp}
\usepackage{xcolor}
\usepackage{float}
\usepackage[hidelinks]{hyperref}
\usepackage{url}
\usepackage{orcidlink}
\def\BibTeX{{\rm B\kern-.05em{\sc i\kern-.025em b}\kern-.08em
    T\kern-.1667em\lower.7ex\hbox{E}\kern-.125emX}}

\usepackage{amsmath}
\usepackage{amsfonts}
\usepackage{amssymb}
\usepackage{nicefrac}
\usepackage{bbm} 
\usepackage{bm} 

\usepackage[all]{xy}

\newcommand{\IR}{\mathbb{R}}		

\renewcommand{\epsilon}{\varepsilon}				
\renewcommand{\theta}{\vartheta}                  	
\renewcommand{\phi}{\varphi}						

\newcommand{\set}[2][]{\left\{ #2 
                        \ifthenelse{\equal{#1}{}}{}{: #1}
                        \right\}}				        

\usepackage[exponent-product = \cdot,separate-uncertainty=true]{siunitx}
\DeclareSIUnit{\mup}{\text{$\mu_0$}}

\usepackage{amsthm}
\theoremstyle{plain} 

\theoremstyle{definition} 

\theoremstyle{remark} 

\usepackage{tikz}
\usepackage{tikz-3dplot}
\usepackage{pgfplots}
\usepackage{pgfplotstable}

\pgfplotsset{compat=1.8}

\usetikzlibrary{shapes,shapes.multipart,positioning,shapes.arrows, arrows.meta, decorations.pathmorphing, calc,fadings}
\usepgfplotslibrary{units,statistics,groupplots}

\pgfplotsset{
    unit markings=slash space,
    /pgfplots/xbar/.style={
    /pgf/bar shift={-0.5*(\numplotsofactualtype*\pgfplotbarwidth + (\numplotsofactualtype-1)*#1) + (.5+\plotnumofactualtype)*\pgfplotbarwidth + \plotnumofactualtype*#1},
    },
}
\pgfplotsset{select coords between index/.style 2 args={
    x filter/.code={
        \ifnum\coordindex<#1\fi
        \ifnum\coordindex>#2\fi
    }
}}

\usepackage{color}
\definecolor{ibilight}{RGB}{193,216,237}
\definecolor{ibidark}{RGB}{0,73,146}	
\definecolor{uke2}{RGB}{170,156,143} 	
\definecolor{uke3}{RGB}{87,87,86}		
\definecolor{ukesec1}{RGB}{255,223,0}	
\definecolor{ukesec2}{RGB}{239,123,5}	
\definecolor{ukesec3}{RGB}{104,195,205}	
\definecolor{ukesec4}{RGB}{138,189,36}	
\definecolor{ukesec5}{RGB}{178,34,41}	
\definecolor{tuhh}{RGB}{45,198,214}     
\definecolor{ibidarkBG}{RGB}{227,229,242}   
\definecolor{uke2BG}{RGB}{233,228,225} 	    
\definecolor{uke3BG}{RGB}{230,231,232}	    
\definecolor{ukesec1BG}{RGB}{255,243,190}   
\definecolor{ukesec2BG}{RGB}{254,232,212}   
\definecolor{ukesec3BG}{RGB}{222,241,241}   
\definecolor{ukesec4BG}{RGB}{233,243,222}   
\definecolor{ukesec5BG}{RGB}{244,230,225}   

\usepackage[capitalise]{cleveref} 

\makeatletter
\newif\iftikz@shading@path

\tikzset{
    shading xsep/.store in=\tikz@pathshadingxsep,
    shading ysep/.store in=\tikz@pathshadingysep,
    shading sep/.style={shading xsep=#1, shading ysep=#1},
    shading sep=0.0cm,
}

\def\tikz@shadepath#1{%
    \iftikz@shading@path%
    \else%
        \tikz@shading@pathtrue%
        \pgfgetpath\tikz@currentshadingpath%
        \begingroup%
            \pgfsys@beginscope
            \tikzset{#1}%
            \xdef\tikz@tmp{\noexpand\def\noexpand\tikz@pathshadingxsep{\tikz@pathshadingxsep}%
                \noexpand\def\noexpand\tikz@pathshadingysep{\tikz@pathshadingysep}}%
            \pgfsys@endscope%
        \endgroup
        \tikz@tmp%
        \pgfextract@process\pgf@shadingpath@southwest{\pgfpointadd{\pgfqpoint{\pgf@pathminx}{\pgf@pathminy}}%
            {\pgfpoint{-\tikz@pathshadingxsep}{-\tikz@pathshadingysep}}}
        \pgfextract@process\pgf@shadingpath@northeast{\pgfpointadd{\pgfqpoint{\pgf@pathmaxx}{\pgf@pathmaxy}}%
            {\pgfpoint{\tikz@pathshadingxsep}{\tikz@pathshadingysep}}}%
        \pgfsetpath\pgfutil@empty%
        \let\tikz@options@saved=\tikz@options%
        \let\tikz@mode@saved=\tikz@mode%
        \let\tikz@options=\pgfutil@empty%
        \let\tikz@mode=\pgfutil@empty%
        \tikz@addoption{%
            \pgfinterruptpath%
            \pgfinterruptpicture%
                \begin{tikzfadingfrompicture}[name=.]
                \pgfscope%
                    \tikzset{shade path/.style=}
                    \path \pgfextra{%
                        \pgfsetpath\tikz@currentshadingpath%
                        \pgf@shadingpath@southwest
                        \expandafter\pgf@protocolsizes{\the\pgf@x}{\the\pgf@y}%
                        \pgf@shadingpath@northeast%
                        \expandafter\pgf@protocolsizes{\the\pgf@x}{\the\pgf@y}%
                        \let\tikz@options=\tikz@options@saved%
                        \let\tikz@mode=\tikz@mode@saved%
                    };
                    \xdef\pgf@shadingboundingbox@southwest{\noexpand\pgfqpoint{\the\pgf@picminx}{\the\pgf@picminy}}%
                    \xdef\pgf@shadingboundingbox@northeast{\noexpand\pgfqpoint{\the\pgf@picmaxx}{\the\pgf@picmaxy}}%
                    \endpgfscope
                \end{tikzfadingfrompicture}%
            \endpgfinterruptpicture%
            \endpgfinterruptpath%
            \pgftransformreset%
            \pgfpathrectanglecorners{\pgf@shadingboundingbox@southwest}{\pgf@shadingboundingbox@northeast}%
            \let\tikz@path@picture=\pgfutil@empty%
            \tikz@mode@fillfalse%
            \tikz@mode@drawfalse%
            \tikz@mode@doublefalse%
            \tikz@mode@clipfalse%
            \tikz@mode@boundaryfalse%
            \tikz@mode@fade@pathfalse%
            \tikz@mode@fade@scopefalse%
            \tikzset{#1}%
            \tikz@mode%
            \def\tikz@path@fading{.}%
            \tikz@mode@fade@pathtrue%
            \tikz@fade@adjustfalse%
            \pgfpointscale{0.5}{\pgfpointadd{\pgf@shadingboundingbox@southwest}{\pgf@shadingboundingbox@northeast}}%
            \edef\tikz@fade@transform{shift={(\the\pgf@x,\the\pgf@y)}}%
            \pgfsetfading{\tikz@path@fading}{\tikz@do@fade@transform}%
            \tikz@mode@fade@pathfalse%
        }%
    \fi%
}
\tikzset{
    shade path/.code={%
        \tikz@addmode{\tikz@shadepath{#1}}%
    }
}
\makeatother 

\begin{document}

\title{Uncertainties of a Spherical Magnetic Field Camera\\
{}
}

\author{\IEEEauthorblockN{Fynn Foerger\IEEEauthorrefmark{1}\IEEEauthorrefmark{2}\IEEEauthorrefmark{3}\orcidlink{0000-0002-3865-4603},
Philip Suskin\IEEEauthorrefmark{1}\IEEEauthorrefmark{2}\orcidlink{0009-0009-0195-4491}, 
Marija Boberg\IEEEauthorrefmark{1}\IEEEauthorrefmark{2}\orcidlink{0000-0003-3419-7481}, Jonas Faltinath\IEEEauthorrefmark{1}\IEEEauthorrefmark{2}\orcidlink{0009-0003-4128-2948}, Tobias Knopp\IEEEauthorrefmark{1}\IEEEauthorrefmark{2}\IEEEauthorrefmark{3}\orcidlink{0000-0002-1589-8517}, \\ and
Martin Möddel\IEEEauthorrefmark{1}\IEEEauthorrefmark{2}\orcidlink{0000-0002-4737-7863}}
\IEEEauthorblockA{
Email: fynn.foerger@tuhh.de\\
\IEEEauthorrefmark{1}Section for Biomedical Imaging, University Medical Center Hamburg-Eppendorf, Hamburg, Germany\\
\IEEEauthorrefmark{2}Institute for Biomedical Imaging, Hamburg University of Technology, Hamburg, Germany\\
\IEEEauthorrefmark{3}Fraunhofer Research Institution for Individualized and Cell-based Medical Engineering IMTE, Lübeck, Germany
}}

\maketitle
\begin{abstract}
Spherical harmonic expansions are well-established tools for estimating magnetic fields from surface measurements and are widely used in applications such as tomographic imaging, geomagnetism, and biomagnetism. Although the mathematical foundations of these expansions are well understood, the impact of real-world imperfections, on the uncertainty of the field model has received little attention.

In this work, we present a systematic uncertainty propagation analysis for a magnetic field camera that estimates the field from surface measurements using a spherical array of Hall magnetometers arranged in a spherical $t$-design. A Monte Carlo–based approach is employed to quantify how sensor-related uncertainties, such as calibration errors and positioning inaccuracies, affect the spatial distribution of the estimated field’s uncertainty. The results offer insights into the robustness of spherical harmonic methods and help identify the dominant sources of uncertainty in practical implementations.
\end{abstract}


\section{Introduction}
Magnetic field measurements play a crucial role in a wide range of industrial and scientific applications. Depending on the application, magnetic field sensing systems must be tailored to capture static or dynamic field distributions with specific requirements regarding precision, dynamic range, and measurement volume.

To measure magnetic fields within a three-dimensional volume, various strategies have been developed. A traditional approach involves moving a single magnetometer through a predefined set of positions, recording the field sequentially at each point\cite{akai_3d_2017}. While this method can offer high spatial resolution, it is highly time-consuming. To overcome these limitations, array-based systems have been proposed that employ multiple sensors distributed throughout or around the measurement volume. By acquiring data simultaneously at all sensor locations, these systems offer significantly increased measurement speed\cite{SCHLAGETER200137,10098773,Nicolas2024fpga}. Depending on the required field range, and sensitivity, a variety of magnetometer technologies can be used, including Hall sensors, fluxgates, and magnetoresistive elements\cite{9296742}.

A particularly effective technique for estimating magnetic fields in a volume is to place sensors on a surface of a predefined volume enclosing the region of interest. In regions with zero current density, the magnetic field satisfies the Laplace equation, allowing each field component to be expressed as a harmonic series. By sampling the field on the surface and solving for the expansion coefficients, the magnetic field throughout the enclosed volume can be accurately reconstructed using a relatively small number of measurement points \cite{dietrich_field_2016,ECCLES1993135,bringout_robust,scheffler_ellipsoidal_2024}.

\begin{figure}
\includegraphics[width=0.24\textwidth]{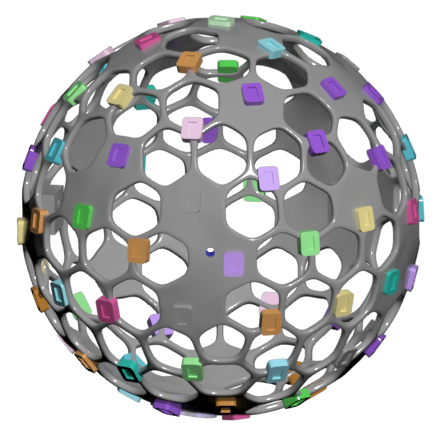}
\includegraphics[width=0.24\textwidth]{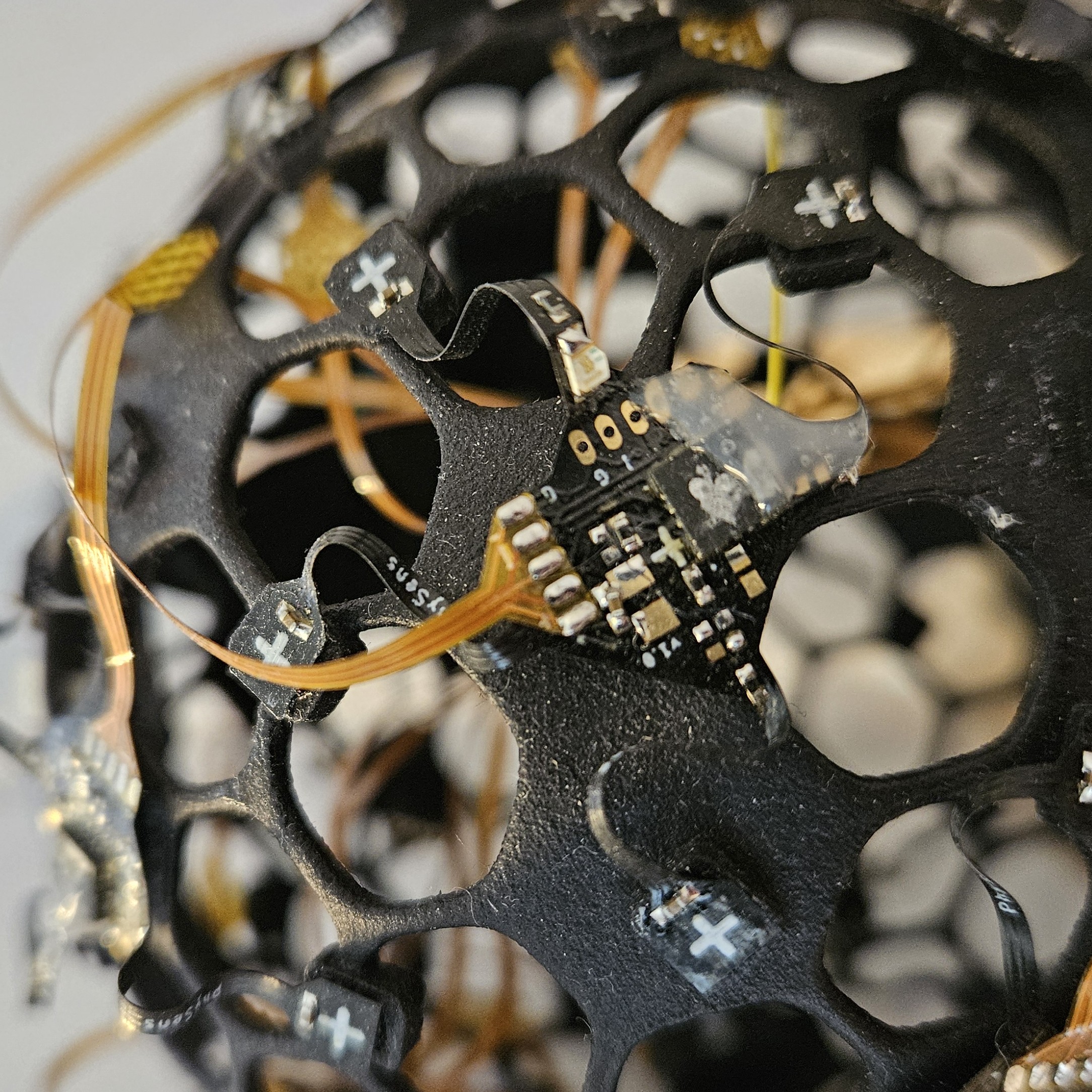}
    \centering
    \caption{Left: CAD rendering of the spherical magnetometer array. Colored brackets hold the individual sensors in place and define their orientation. The hexagonal openings enable convenient routing of the flexible circuit connections without compromising the mechanical stability. Right: Detailed view of a flexible printed circuit board with integrated Hall sensors. Reprinted from \cite{foerger2025realtime3dmagneticfield}.}
    \label{fig:sph_render}
\end{figure}

In this work, we build upon a previously developed spherical magnetometer array, which provides surface measurements used to construct a surrogate model of the magnetic field via a spherical harmonic expansion~\cite{foerger_flexible_2023}. The array consists of $N = 86$ Hall sensors placed on a 3D-printed sphere with a diameter of \SI{9}{\cm}, following a spherical $t$-design with $t = 12$~\cite{Boberg_Knopp_Möddel_2025,Beentjes2016QUADRATUREOA}. It enables the simultaneous acquisition of full 3D vector field data at a temporal resolution of \SI{10}{\Hz}, with field strengths covered ranging from \SI{\pm 266}{\milli\tesla}. The reconstructed field is represented by a sixth-degree spherical harmonic expansion, resulting in a compact and smooth volumetric field description. A picture and CAD rendering of the sensor is shown in Fig.~\ref{fig:sph_render}.

While previous work focused on system architecture and calibration method, this study investigates how real-world uncertainties, such as sensor calibration uncertainties and placement inaccuracies, propagate into the reconstructed field. We present a systematic analysis of key uncertainty sources and quantify their individual and combined influence on the reconstruction accuracy.

\section{Methods}
\subsection{Field Expansion}
The field expansion for a discrete measurement on the surface of a Ball $\mathbb{B}_R$ with radius $R$ according to a $t$-design can be expressed as
\begin{align}
    B^i(\bm{r}) \approx \sum_{l=0}^{\lfloor t/2 \rfloor} \sum_{m=-l}^l \gamma_{lm}^i Z_l^m(\bm{r}) \quad \forall \bm{r} \in \mathbb{B}_R,
    \label{eq:expansionShort}
\end{align}
for each field component $i \in \{x,y,z\}$, where \(\gamma_{lm}^i \in \mathbb{R}\) are the expansion coefficients, and \(Z_l^m(\bm{r})\) are the solid spherical harmonics~\cite{Boberg_Knopp_Möddel_2025}.

With the $t$-design sampling points \(\{\bm{r}_1, \dots, \bm r_N\} \subseteq \partial\mathbb{B}_R\) the expansion coefficients are given by
\begin{align}
    \gamma_{lm}^i = \frac{2l+1}{R^l N} \sum_{k=1}^N B^i_k Z_l^m\left(\frac{\bm{r}_k}{R}\right),
    \label{eq:coeffsSum}
\end{align}
where \(B^i_k\) are the measured components of the magnetic field at $\{\bm{r}_1, \dots, \bm r_N\}$~\cite{foerger2025realtime3dmagneticfield}. For a more detailed analysis, see  \cite{Boberg_Knopp_Möddel_2025}.

\subsection{Uncertainty-Prone Variables}
If the magnetic field can be accurately described by the truncated expansion in \eqref{eq:expansionShort}, then any reconstruction uncertainty in the field \(B^i\) originates from uncertainties in the expansion coefficients \(\gamma_{lm}^i\) as defined in \eqref{eq:coeffsSum}. These coefficients are influenced by several sources of measurement uncertainty. 

In our setup, we apply a linear calibration model to the measured field vectors, given by $\bm B_k = \bm R_k \bm b_k + \bm O_k$, where $\bm R_k \in \IR^{3\times3}$ represents the rotation and scaling matrix, $\bm O_k \in \IR^3$ the offset correction, and $\bm b_k \in \IR^3$ the raw sensor reading~\cite{foerger2025realtime3dmagneticfield}. Substituting this into the expression for the expansion coefficients yields
\begin{align}
    \gamma_{lm}^i = \frac{2l+1}{R^l N} 
    \sum_{k=1}^N 
    (\bm R_k \bm b_k+\bm O_k)^i 
    Z_l^m\left(
      \frac{
        \bm{r}_k
      }{R}
    \right).
    \label{eq:coeffsSumWithCalib}
\end{align}

The key variables contributing to uncertainty in this equation are:

\begin{itemize}
    \item $\bm b_k$: The raw sensor readings are subject to various sensor-specific uncertainties. For the TMAG5273x2 (Texas Instruments) used in our field camera, this includes a sensitivity drift of up to 5\%, an offset drift of approximately \SI{3}{\micro\tesla}, and a 1 sigma magnetic noise of \SI{24}{\micro\tesla}.
    
    \item $\bm R_k$, $\bm O_k$: The rotation matrices and offset vectors are determined by exposing each sensor to six orthogonal magnetic fields of \SI{20}{\milli\tesla}, generated by a dedicated calibration setup~\cite{foerger2025realtime3dmagneticfield}. A total of 3000 measurements were acquired, and the calibration parameters were obtained by solving an optimization problem~\cite{foerger2025realtime3dmagneticfield}. Uncertainties in $\bm R_k$ and $\bm O_k$ arise primarily from spatial inhomogeneities in the calibration field and sensor noise. The field exhibits a maximum norm inhomogeneity of $1.1 \cdot 10^{-3}$ and a maximum undesired orthogonal component of $1.2\%$. Additionally, the Earth's magnetic field was not compensated during calibration, introducing an offset uncertainty of approximately \SI{50}{\micro\tesla}.
    
    \item $\bm r_k$: Sensor positions are inferred from the rotation matrices $\bm R_k$ using an optimization procedure. Each sensor is assumed to be mounted tangentially to the surface of the sphere, allowing the position to be deduced from its local z axis~\cite{foerger2025realtime3dmagneticfield}. In practice, however, the manual placement introduces an estimated angular uncertainty of about 2°. However, since $\bm R_k$ is calculated from the calibration measurements, the inhomogeneity of the calibration field further contributes to the uncertainty.
    
    \item $R$: The sensors are mounted on a 3D-printed spherical support structure. Due to limited printing accuracy and possible thermal deformations, the actual radius may deviate from the design value. To account for this, we treat \(R\) as a free parameter within a tolerance of \SI{0.5}{\mm}. The optimal radius is determined in a subsequent fitting procedure by aligning the reconstructed field with an independent reference measurement. For this reason, we neglect the uncertainty in \(R\) in the following analysis.
\end{itemize}

\subsection{Uncertainty Propagation}
\begin{figure*}[!ht]
    \centering
    \includegraphics[width=\linewidth]{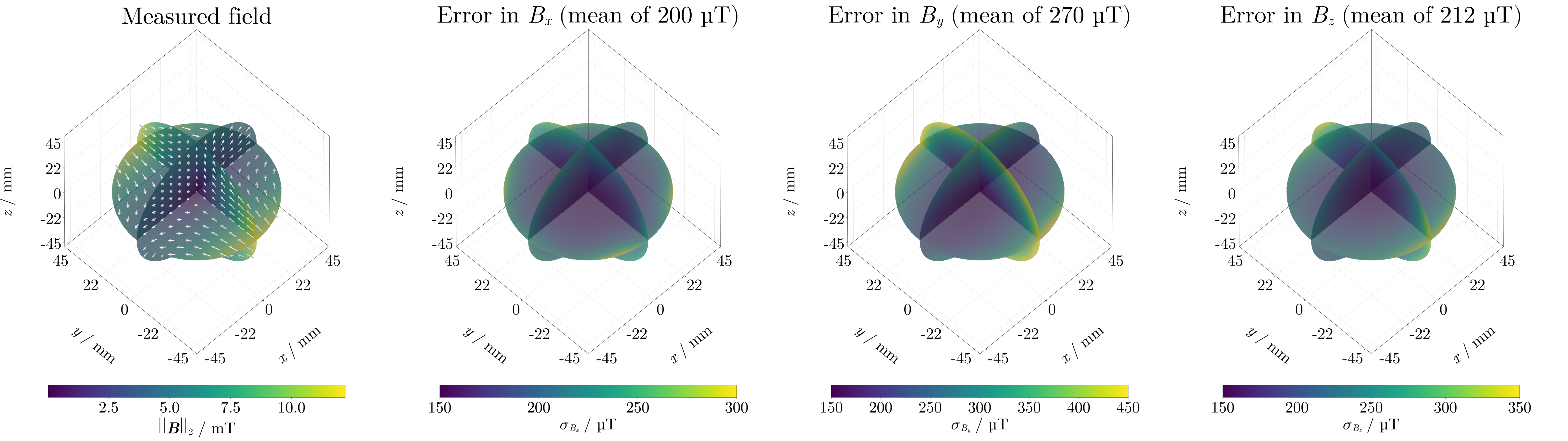}
    \caption{First: The measured field is a static gradient field with a field-free point at the center, exhibiting a gradient strength of \SI{0.22}{\tesla\per\meter} along the $y$-axis and \SI{0.11}{\tesla\per\meter} along the $x$- and $z$-axes.
    Second to Fourth: Propagated absolute uncertainties for the individual magnetic field components. Please note that the color scales differ between plots and are scaled to the respective plot's maximum and minimum values.}
    \label{fig:ResultsMakie}
\end{figure*}
The coefficients \(\gamma_{lm}^i\) include variables derived from optimization problems that depend on uncertain input parameters. For these cases, the derivatives required for the uncertainty propagation cannot be analytically determined. Therefore, we adopt a Monte Carlo–based approach to estimate the covariance matrix $\Sigma_\Gamma$ of the coefficients \(\gamma_{lm}^i\), containing
$\sigma_{\gamma^i_{\tilde{l}\tilde{m}}\gamma^i_{lm\vphantom{\tilde{l}}}}^2$,
using the Julia package \texttt{MonteCarloMeasurements.jl}~\cite{carlson2020montecarlomeasurementsjlnonlinearpropagationarbitrary}. We simulate each uncertain variable using 10000 particles to emulate the effect on the coefficients of the expansion.

Let \(\bm \Gamma\) be the vector containing all the coefficients $\gamma^i_{lm}$, the spatially dependent variance \(\sigma_{B^i}^2(\bm{r})\) of the reconstructed field component \(B^i\) inside the sphere can be expressed as
\begin{align}
    \begin{split}
    \sigma_{B^i}^2(\bm{r}) &= \left(\nabla_{\bm \Gamma} B^i\left(\bm r\right)\right)^T \Sigma_{\bm \Gamma} \left(\nabla_{\bm \Gamma} B^i \left( \bm r\right)\right) \\
    &=\sum_{\tilde{l}=0}^6\sum_{\tilde{m}=-\tilde{l}}^{\tilde{l}}\sum_{l=0}^6\sum_{m=-l}^l\sigma_{\gamma^i_{\tilde{l}\tilde{m}}\gamma^i_{lm\vphantom{\tilde{l}}}}^2 Z_{\tilde{m}}^{\tilde{l}}(\bm r)Z_{m}^{l}(\bm r) .
    \end{split}
\end{align}

The following assumptions are made for modeling the input uncertainties:

\begin{itemize}
    \item \textbf{\(\bm b_k\)}: Sensor reading uncertainties are modeled as uncorrelated across sensors. The total variance is assumed to follow a normal distribution with variance \(\left(0.05 \cdot |\bm b_k|\right)^2 + (\SI{3}{\micro\tesla})^2 + (\SI{24}{\micro\tesla})^2\), accounting for sensitivity drift, offset drift, and sensor noise.
    
    \item \textbf{\(\bm R_k\), \(\bm O_k\)}: Three main sources of uncertainty are included.  First, spatial inhomogeneities in the calibration field are modeled by randomly perturbing the field vectors with small rotations and scaling variations. These perturbations follow normal distributions based on the previously mentioned homogeneity characterization values. The field within a solenoid naturally has a certain structure. However, since the measuring device was held in six different orientations in the calibration coil for each of the calibration measurements, we assume that these uncertainties are uncorrelated. Second, the Earth's magnetic field is modeled as an unknown offset vector with uniformly distributed random orientation. 
    Third, the measured calibration field vectors, though based on 3000 samples, still exhibit a small uncertainty, which is estimated from the measurements.
    
    \item \textbf{\(\bm r_k\)}: Since the sensor positions \(\bm r_k\) are determined through an optimization procedure based on \(\bm R_k\), we reuse the previously sampled distribution of \(\bm R_k\) to infer the corresponding distribution of positions. This approach inherently accounts for correlations between orientation and position uncertainties.
\end{itemize}

To assess the dominant sources of uncertainty, we also evaluated the effect of each uncertainty individually in isolated Monte Carlo runs.

\subsection{Field Under Consideration}

Since the absolute field uncertainty depends on the field magnitude itself, we evaluate the reconstruction uncertainty using a specific example field. For this purpose, we consider a static gradient field generated by a head-sized magnetic particle imaging scanner. The field features a well-defined field-free point and a gradient strength of \SI{0.22}{\tesla\per\meter} in the $y$-direction and \SI{0.11}{\tesla\per\meter} in the $x$- and $z$-directions as shown in Fig.~\ref{fig:ResultsMakie} on the left~\cite{thieben_system_2024}.

\section{Results}

The propagated absolute uncertainties $\sigma_{B^i}$ for each field component within the sensor sphere are shown in the three plots on the right of Fig.~\ref{fig:ResultsMakie}. The sampled uncertainties are smallest at the center of the sphere, where the measured magnetic field is weakest. The largest uncertainties occur near the outer boundary of the sphere. However, since the uncertainty gradients are significantly smaller than the gradients of the field itself, the relative uncertainty decreases from the center, toward the outer regions. The highest absolute uncertainties are observed in the $y$-component, which reflects the fact that the original field exhibits its highest values in the $y$-direction. 
However, the average uncertainty for the $x$, $y$ and $z$ component inside the sphere is \SI{200}{\micro\tesla}, \SI{270}{\micro\tesla} and \SI{212}{\micro\tesla}.

The Monte Carlo simulations with isolated uncertainty sources reveal that the most significant contribution stems from inhomogeneities in the calibration field combined with the unaccounted Earth's magnetic field. This source results in an average uncertainty of the norm of the field of \SI{308}{\micro\tesla} within the sphere.
The second largest contribution arises from sensor positioning inaccuracies, leading to a mean uncertainty of \SI{221}{\micro\tesla}.
The smallest impact is caused by sensor-specific effects such as drift and noise, contributing \SI{138}{\micro\tesla} on average.

\section{Discussion and Conclusion}
We presented an uncertainty propagation analysis for a surrogate model of the magnetic field, constructed via a spherical harmonic expansion based on measurements from a spherical Hall sensor array. In principle, the field determination method is very robust against uncorrelated measurement uncertainties of the sensors, as the field coefficients are always calculated globally from all measured values. The largest contribution to uncertainty stems from spatial inhomogeneities in the calibration field. This underscores the importance of a precise, homogeneous calibration to achieve reliable field measurements in practical applications, and should be a key focus. 
Although sensor-specific effects such as sensitivity drift are larger in magnitude, their influence averages out under the assumption of uncorrelated uncertainties. However, this assumption breaks down if global uncertainty sources, such as temperature drift, are not compensated for and affect multiple sensors similarly.

The field under investigation was well approximated by an expansion with a polynomial degree of 6, but for more complex fields, truncation uncertainties must be considered. Moreover, the analysis assumes a perfectly spherical geometry, which may not hold in practice due to manufacturing or thermal effects. Such systematic deviations could distort the reconstructed field and should be addressed in future uncertainty models.

\bibliographystyle{IEEEtran}

\bibliography{ref}

\begin{thebibliography}{10}
\providecommand{\url}[1]{#1}
\csname url@samestyle\endcsname
\providecommand{\newblock}{\relax}
\providecommand{\bibinfo}[2]{#2}
\providecommand{\BIBentrySTDinterwordspacing}{\spaceskip=0pt\relax}
\providecommand{\BIBentryALTinterwordstretchfactor}{4}
\providecommand{\BIBentryALTinterwordspacing}{\spaceskip=\fontdimen2\font plus
\BIBentryALTinterwordstretchfactor\fontdimen3\font minus \fontdimen4\font\relax}
\providecommand{\BIBforeignlanguage}[2]{{%
\expandafter\ifx\csname l@#1\endcsname\relax
\typeout{** WARNING: IEEEtran.bst: No hyphenation pattern has been}%
\typeout{** loaded for the language `#1'. Using the pattern for}%
\typeout{** the default language instead.}%
\else
\language=\csname l@#1\endcsname
\fi
#2}}
\providecommand{\BIBdecl}{\relax}
\BIBdecl

\bibitem{akai_3d_2017}
\BIBentryALTinterwordspacing
N.~Akai and K.~Ozaki, ``\BIBforeignlanguage{en}{{3D} magnetic field mapping in large-scale indoor environment using measurement robot and {Gaussian} processes},'' in \emph{\BIBforeignlanguage{en}{2017 {International} {Conference} on {Indoor} {Positioning} and {Indoor} {Navigation} ({IPIN})}}.\hskip 1em plus 0.5em minus 0.4em\relax Sapporo: IEEE, Sep. 2017, pp. 1--7. [Online]. Available: \url{http://ieeexplore.ieee.org/document/8115960/}
\BIBentrySTDinterwordspacing

\bibitem{SCHLAGETER200137}
\BIBentryALTinterwordspacing
V.~Schlageter, P.-A. Besse, R.~Popovic, and P.~Kucera, ``Tracking system with five degrees of freedom using a 2d-array of hall sensors and a permanent magnet,'' \emph{Sensors and Actuators A: Physical}, vol.~92, no.~1, pp. 37--42, 2001, selected Papers for Eurosensors XIV. [Online]. Available: \url{https://www.sciencedirect.com/science/article/pii/S0924424701005374}
\BIBentrySTDinterwordspacing

\bibitem{10098773}
\BIBentryALTinterwordspacing
C.~Vergne, J.~Inácio, T.~Quirin, D.~Sargent, M.~Madec, and J.~Pascal, ``Tracking of a magnetically navigated millirobot with a magnetic-field camera,'' \emph{IEEE Sensors Journal}, vol.~24, no.~6, pp. 7336--7344, 2024. [Online]. Available: \url{https://ieeexplore.ieee.org/document/10098773}
\BIBentrySTDinterwordspacing

\bibitem{Nicolas2024fpga}
\BIBentryALTinterwordspacing
H.~Nicolas, T.~Quirin, and J.~Pascal, ``{FPGA-Based Magnetic Field Camera for Dynamic Magnetic Field Mapping},'' \emph{IEEE Sensors Letters}, vol.~8, no.~5, pp. 1--4, 2024. [Online]. Available: \url{https://ieeexplore.ieee.org/document/10502211}
\BIBentrySTDinterwordspacing

\bibitem{9296742}
\BIBentryALTinterwordspacing
N.~Hadjigeorgiou, K.~Asimakopoulos, K.~Papafotis, and P.~P. Sotiriadis, ``Vector magnetic field sensors: Operating principles, calibration, and applications,'' \emph{IEEE Sensors Journal}, vol.~21, no.~11, pp. 12\,531--12\,544, 2021. [Online]. Available: \url{https://ieeexplore.ieee.org/document/9296742}
\BIBentrySTDinterwordspacing

\bibitem{dietrich_field_2016}
\BIBentryALTinterwordspacing
B.~E. Dietrich, D.~O. Brunner, B.~J. Wilm, C.~Barmet, S.~Gross, L.~Kasper, M.~Haeberlin, T.~Schmid, S.~J. Vannesjo, and K.~P. Pruessmann, ``\BIBforeignlanguage{en}{A field camera for {MR} sequence monitoring and system analysis},'' \emph{\BIBforeignlanguage{en}{Magnetic Resonance in Medicine}}, vol.~75, no.~4, pp. 1831--1840, Apr. 2016. [Online]. Available: \url{https://onlinelibrary.wiley.com/doi/10.1002/mrm.25770}
\BIBentrySTDinterwordspacing

\bibitem{ECCLES1993135}
\BIBentryALTinterwordspacing
C.~Eccles, S.~Crozier, M.~Westphal, and D.~Doddrell, ``Temporal spherical-harmonic expansion and compensation of eddy-current fields produced by gradient pulses,'' \emph{Journal of Magnetic Resonance, Series A}, vol. 103, no.~2, pp. 135--141, 1993. [Online]. Available: \url{https://www.sciencedirect.com/science/article/pii/S1064185883711447}
\BIBentrySTDinterwordspacing

\bibitem{bringout_robust}
\BIBentryALTinterwordspacing
G.~Bringout and T.~M. Buzug, ``\BIBforeignlanguage{en}{A robust and compact representation for magnetic ﬁelds in magnetic particle imaging},'' \emph{\BIBforeignlanguage{en}{Biomed Tech}}, vol.~59, pp. 978--1, 2014. [Online]. Available: \url{http://www.gael-bringout.com/public/Bringout%202014%20-%20a%20robust%20and%20compact%20representation%20for%20magnetics%20fields%20in%20magnetic%20particle%20imaging.pdf}
\BIBentrySTDinterwordspacing

\bibitem{scheffler_ellipsoidal_2024}
\BIBentryALTinterwordspacing
K.~Scheffler, L.~Meyn, F.~Foerger, M.~Boberg, M.~Martin, and T.~Knopp, ``\BIBforeignlanguage{en}{Ellipsoidal {Harmonic} {Expansions} for {Efficient} {Approximation} of {Magnetic} {Fields} in {Medical} {Imaging}},'' \emph{\BIBforeignlanguage{en}{International Journal on Magnetic Particle Imaging IJMPI}}, vol.~10, no. 1 Suppl 1, Mar. 2024, publisher: International Journal on Magnetic Particle Imaging IJMPI. [Online]. Available: \url{https://www.journal.iwmpi.org/index.php/iwmpi/article/view/725}
\BIBentrySTDinterwordspacing

\bibitem{foerger_flexible_2023}
\BIBentryALTinterwordspacing
F.~Foerger, N.~Hackelberg, M.~Boberg, J.-P. Scheel, F.~Thieben, L.~Mirzojan, F.~Mohn, M.~Möddel, M.~Graeser, and T.~Knopp, ``\BIBforeignlanguage{en}{Flexible {Selection} {Field} {Generation} using {Iron} {Core} {Coil} {Arrays}},'' \emph{\BIBforeignlanguage{en}{International Journal on Magnetic Particle Imaging IJMPI}}, p. Vol 9 No 1 Suppl 1 (2023), Mar. 2023, publisher: International Journal on Magnetic Particle Imaging IJMPI. [Online]. Available: \url{https://journal.iwmpi.org/index.php/iwmpi/article/view/624}
\BIBentrySTDinterwordspacing

\bibitem{Boberg_Knopp_Möddel_2025}
\BIBentryALTinterwordspacing
M.~Boberg, T.~Knopp, and M.~Möddel, ``Unique compact representation of magnetic fields using truncated solid harmonic expansions,'' \emph{European Journal of Applied Mathematics}, p. 1–28, 2025. [Online]. Available: \url{https://doi.org/10.1017/S0956792524000883}
\BIBentrySTDinterwordspacing

\bibitem{Beentjes2016QUADRATUREOA}
\BIBentryALTinterwordspacing
C.~H.~L. Beentjes, ``Quadrature on a spherical surface,'' 2016. [Online]. Available: \url{https://api.semanticscholar.org/CorpusID:19681887}
\BIBentrySTDinterwordspacing

\bibitem{foerger2025realtime3dmagneticfield}
\BIBentryALTinterwordspacing
F.~Foerger, M.~Boberg, N.~Hackelberg, P.~Heinisch, K.~Ostaszewski, J.~Faltinath, F.~Thieben, F.~Mohn, P.~Jürß, M.~Möddel, and T.~Knopp, ``Real-time 3d magnetic field camera for a spherical volume,'' 2025. [Online]. Available: \url{https://arxiv.org/abs/2503.04391}
\BIBentrySTDinterwordspacing

\bibitem{carlson2020montecarlomeasurementsjlnonlinearpropagationarbitrary}
\BIBentryALTinterwordspacing
F.~B. Carlson, ``{MonteCarloMeasurements.jl: Nonlinear Propagation of Arbitrary Multivariate Distributions by means of Method Overloading},'' 2020. [Online]. Available: \url{https://arxiv.org/abs/2001.07625}
\BIBentrySTDinterwordspacing

\bibitem{thieben_system_2024}
\BIBentryALTinterwordspacing
F.~Thieben, F.~Foerger, F.~Mohn, N.~Hackelberg, M.~Boberg, J.-P. Scheel, M.~Möddel, M.~Graeser, and T.~Knopp, ``\BIBforeignlanguage{en}{System characterization of a human-sized {3D} real-time magnetic particle imaging scanner for cerebral applications},'' \emph{\BIBforeignlanguage{en}{Communications Engineering}}, vol.~3, no.~1, p.~47, Mar. 2024. [Online]. Available: \url{https://www.nature.com/articles/s44172-024-00192-6}
\BIBentrySTDinterwordspacing

\end{thebibliography}

\end{document}